\documentclass[]{spie}

\usepackage{pbox}
\usepackage{multirow}
\usepackage{amsmath,amsfonts,amssymb}
\usepackage{graphicx}
\usepackage{floatrow}
\usepackage[export]{adjustbox}
\graphicspath{{.}} 
\usepackage{subcaption}
\usepackage{sidecap}
\usepackage[colorlinks=true, allcolors=blue]{hyperref}
\usepackage{enumerate}
\usepackage[flushleft]{threeparttable}
 
\title{Panoramic optical and near-infrared SETI instrument: optical and structural design concepts} 

\author{
J\'er\^ome Maire\supit{a}, Shelley A. Wright\supit{a,b}, Maren Cosens\supit{a,b}, Franklin P. Antonio\supit{c}, \\Michael L. Aronson\supit{d}, 
Samuel A. Chaim-Weismann\supit{e}, Frank D. Drake\supit{f}, Paul Horowitz\supit{g},\\ Andrew W. Howard\supit{h}, Geoffrey W. Marcy\supit{e}, Rick Raffanti\supit{i}, Andrew P. V. Siemion\supit{e,f,j,k}, Remington P. S. Stone\supit{l}, Richard R. Treffers\supit{m}, Avinash Uttamchandani\supit{n}, Dan Werthimer\supit{e,o}
\skiplinehalf
\supit{a}Center for Astrophysics \& Space Sciences, University of California San Diego, USA; \\
\supit{b}Department of Physics, University of California San Diego, USA; \\
\supit{c}Qualcomm, San Diego, CA, USA; \\
\supit{d}Electronics Packaging Man, San Diego, CA, USA; \\
\supit{e}Department of Astronomy, University of California Berkeley, CA, USA; \\
\supit{f}SETI Institute, Mountain View, CA, USA; \\
\supit{g}Department of Physics, Harvard University, Cambridge, MA, USA; \\
\supit{h}Astronomy Department, California Institute of Technology, USA; \\
\supit{i}Techne Instruments, Oakland, CA, USA; \\
\supit{j}Radboud University, Nijmegen, Netherlands; \\
\supit{k}Institute of Space Sciences and Astronomy, University of Malta;\\
\supit{l}University of California Observatories, Lick Observatory, USA; \\
\supit{m}Starman Systems, Alamo, CA, USA; \\
\supit{n}Nonholonomy, LLC, Cambridge, USA; \\
\supit{o}Space Sciences Laboratory, University of California Berkeley, CA, USA
}

\authorinfo{Further author information: (Send correspondence to J.M.)\\J.M.: E-mail: jmaire@ucsd.edu}

\begin{document} 
\maketitle 

\begin{abstract}
We propose a novel instrument design to greatly expand the current optical and near-infrared SETI search parameter space by monitoring the entire observable sky during all observable time. This instrument is aimed to search for technosignatures by means of detecting nano- to micro-second light pulses that could have been emitted, for instance, for the purpose of interstellar communications or energy transfer. We present an instrument conceptual design based upon an assembly of 198 refracting 0.5-m telescopes tessellating two geodesic domes. This design produces a regular layout of hexagonal collecting apertures that optimizes the instrument footprint, aperture diameter, instrument sensitivity and total field-of-view coverage. We also present the optical performance of some Fresnel lenses envisaged to develop a dedicated panoramic SETI (PANOSETI) observatory that will dramatically increase sky-area searched (pi steradians per dome), wavelength range covered, number of stellar systems observed, interstellar space examined and duration of time monitored with respect to previous optical and near-infrared technosignature finders.
\end{abstract}


\section{INTRODUCTION}\label{sec:intro}
The SETI (Search for ExtraTerrestrial Intelligence) search parameter space has been considerably expanded since the first dedicated observations in 1960 using the 26-m Tatel radio telescope\cite{Drake1961}. Sensitivity, bandwidth, spectral resolution and data handling capacities have increased over the years, but rare efforts have been made to design instruments with wide instantaneous field-of-view (FoV), i.e. the region of the sky that can be examined in a single observation. While the fraction of Sun-like stars hosting Earth-size planets in their habitable zone may be relatively high in our Galaxy ($22\pm 4$\%\cite{Petigura2013}), the fraction of these planets on which life can develop civilizations exposing signs of technology remains totally unknown; technosignatures could be rare and transient. Wide-field instrument designs are hence of crucial importance for surveying large areas of sky. 
Ultimately, SETI instruments capable of performing all-sky all-time observations with sufficient sensitivity would provide the largest possible angular and temporal coverage, increasing the probability of detecting a transient phenomena coming from an unknown location. 

Single-aperture wide-field astronomical instruments usually have complex optical designs aimed at minimizing aberrations and maintaining image quality over large fields of view, i.e. a few square degrees. Wide-field large-aperture telescopes made for large surveys such as QUEST\cite{Djorgovski2008}, Pan-STARRS\cite{Chambers2016}, SDSS\cite{Gunn2006}, or the future LSST\cite{Angel2001} have instantaneous FoV of 4, 3, 2.5 and 3.5 degrees respectively\cite{Ackermann2010}. FoV of conventional radio telescopes, limited to the beam area ($\sim\lambda/D$) of a single-dish, can be significantly enlarged if equipped with phased array feed receivers, such as ASKAP\cite{Heywood2016} ( 30 square degrees FoV at 1.4GHz) and WSRT/Apertif \cite{Oosterloo2010} (8 degrees across). Assemblies of single-aperture telescopes capable of observing different parts of the sky are still needed to survey the entire sky quickly and repeatedly. Optical large-aperture telescope arrays have been deployed for this purpose, such as the Fly's Eye Cosmic Ray Detector\cite{Baltrusaitis1985} or the Telescope Array\cite{Tokuno2012}. In radio astronomy, development of low-frequency wide-field telescopes has been made possible with aperture arrays constituted of large assemblies of small, fixed antennas, such as MWA\cite{Tingay2013} (with a FoV of 30 degrees across, at a resolution of several arcminutes), LWA\cite{Ellingson2009} (8 degrees across at a resolution of 8 arcsec) and LOFAR \cite{vanHaarlem2013} (FoV of 2$\pi$ steradians in its low-sensitivity survey mode). 
Radio interferometer arrays can also be configured for non-interferometric observations, where each dish examines an unique part of the sky, such as the Allen Telescope Array in Fly's Eye mode\cite{Siemion2012} (200 sq.deg at 1.4GHz).

Transit observing strategies have been adopted to perform the first optical SETI all-sky surveys\cite{Howard2000, Howard2007} with 0.32 sq.deg of instantaneous FoV covering the sky in 150 clear nights. We propose in this paper a novel instrument designed to greatly expand the current optical and near-infrared SETI phase space by monitoring the entire observable sky during all observable time, in order to search for nano- to micro-second pulsed light signals. We present an instrument conceptual design, based upon an assembly of 198 Fresnel-lens telescopes tessellating two geodesic domes which could be used to search for optical technosignatures. The Pulsed All-sky Near-infrared and Optical SETI instrument project (PANOSETI) is dedicated to the search for technosignatures by means of detecting short ($<$1,000ns) visible and near-infrared pulse emissions with a high sensitivity ($\sim$ 25-50 photons per pulse per aperture). Equipped with fast ($>$200MHz) low-noise visible and near-infrared detector arrays \cite{Wright2018}, each part of the sky will be observed simultaneously from two locations to detect coincident signals and minimize false alarms generated by various sources of noise. With an instantaneous field-of-view covering 20.47\% of the sky (8,445 square degrees), the PANOSETI observatory, further described in Wright et al.\cite{Wright2018} and Cosens et al.\cite{Cosens2018}, is aimed to perform a high-sensitivity panoramic search for technosignatures from the Northern Hemisphere. 

\section{STRUCTURAL DESIGN}\label{sec:struc}
The compact dome structural design proposed for PANOSETI, generated from the triangular subdivision of a truncated spherical icosahedron, produces a regular layout of hexagonal collecting apertures with minimal overlap or gaps. It also optimizes the instrument footprint, and maximizes aperture diameter, instrument sensitivity and total field-of-view coverage. This design also minimizes the amount of material required for building the frame structure supporting individual telescope modules and assembly volume in comparison to any other design.

\subsection{ALL-SKY DESIGN CONSIDERATIONS} \label{sec:structural-allsky}
Unless observing from space, monitoring the full $4\pi$-sr sky, all the time can only be done from multiple observing sites, and only intermittently due to daytime optical observation interruptions. Since it is common practice to avoid ground-based optical observations through high air-masses ($>2$, i.e. elevations lower than $30^\circ$) due to higher sky-background induced by distant city lights and higher light extinction, the maximal simultaneous FoV per observatory is then limited to $\pi$-sr ($\sim 10,000$ sq.deg.) during clear-sky night time. For instance, such a wide instantaneous FoV would require 100 individual telescopes, each with a 10x10 degree FoV. Such large FoV per telescope could be obtained at low angular resolution, with small-aperture ($<$10cm typically) wide-field cameras\cite{Law2016, Meszaros2016 } or some large-aperture telescopes with a degraded off-axis image quality. Because the primary goal of SETI experiments is to determine the prevalence of intelligent (technologically capable) life in the universe, a fine localization of its source could be considered as a secondary objective; optical SETI experiments do not necessarily require high-angular resolution, but large apertures for sensitivity.

The spatial arrangement of numerous collecting apertures viewing different parts of the sky can impact performance, complexity and cost of the overall instrument. For instance, separation between on-sky projected FoV should be kept as constant as possible so as to avoid frequent monitoring and correction of the pointing directions. An array configuration of separated telescopes would have independent mount deformations (thermal expansion, wind load, soil movements) that would need frequent telescope FoV coordinate measurements and corrections. The extremely short integration time needed for pulsed optical SETI (less than one micro-second) removes the need for tracking components and moving parts. A fixed structure, supporting all the telescopes, is then appropriate to maintain relative mis-alignments under acceptable levels requiring less frequent calibrations. Mosaic arrangements on a spherical frame have been considered for packing apertures of all-sky surveying instruments such as the Cornell Cosmic House\cite{Bunner1967,Bunner1968}, EvryScope\cite{Law2016}, or Fly's Eye camera system\cite{Pal2013}. A spherical frame is the most compact structure that could be designed for carrying an assembly of multiple-apertures intended to survey the full sky. This compactness optimizes the total footprint and height of the all-sky observatory. It can also minimize the number of mount elements since adjacent apertures can have shared mount support. A compact spherical assembly can also reduce the number of required telescope shelters to a single spherical enclosure having the smallest possible surface area.

In optical pulse SETI experiments, false alarm rates can be minimized by using several detectors covering the same portion of the sky in search for a coincident detection of pulses\cite{Horowitz2001,Werthimer2001,Howard2004,Wright2014,Maire2014,Maire2016}. In order to minimize false alarms generated by “background” events induced by cascades of ionized particles and electromagnetic radiation (Cherenkov photons) from the propagation of gamma-rays and cosmic rays in the Earth's atmosphere, it is preferable to have widely separated detectors working in coincidence. Since Cherenkov photons are spread almost uniformly over a circular area about 250 meters in diameter\cite{Covault2001}, detectors working in coincidence should be separated by at least this distance. 

A spherical frame structure used to support telescope modules could be made of straight bars for ease of manufacturing. Long, unsupported members should not be used due to increasing bending from applied shear forces. Gravitational loads, such as telescopes, will cause less pronounced bending of frame members if applied at or near the joints. The frame structure should be designed in a way that members should not obscure any part of the module apertures. If the modules are all identical, tessellating the hemisphere with regular polygons will ensure optimal use of member material and minimize the size of the frame structure. 

To design and test the geometric layout of apertures, members are first simplified as line segments connected at points (representing joints). Optimally, we would like to divide the sphere into a grid of regular polygons, evenly distributed. Variations within the grid (chords and areas) should be minimized and some metrics should be defined to compare one subdivision method with another (Sect.\ref{sec:structural-tessell}). To subdivide the sphere evenly, one could begin with the icosahedron (12 vertices, 20 equilateral triangular faces, 30 edges) which has the highest number of identical faces of any regular polyhedron, subdividing the sphere into the most sections of equal areas. The icosahedron vertices are already evenly distributed on its circumscribing sphere and we can use this to define great circles and reference points for further subdivisions. Any grid we develop on one face of the spherical icosahedron can be replicated to cover others, thus covering the entire sphere with a pattern that has no overlap or gaps.

\begin{figure}[htb]
\includegraphics[width=0.46\textwidth]{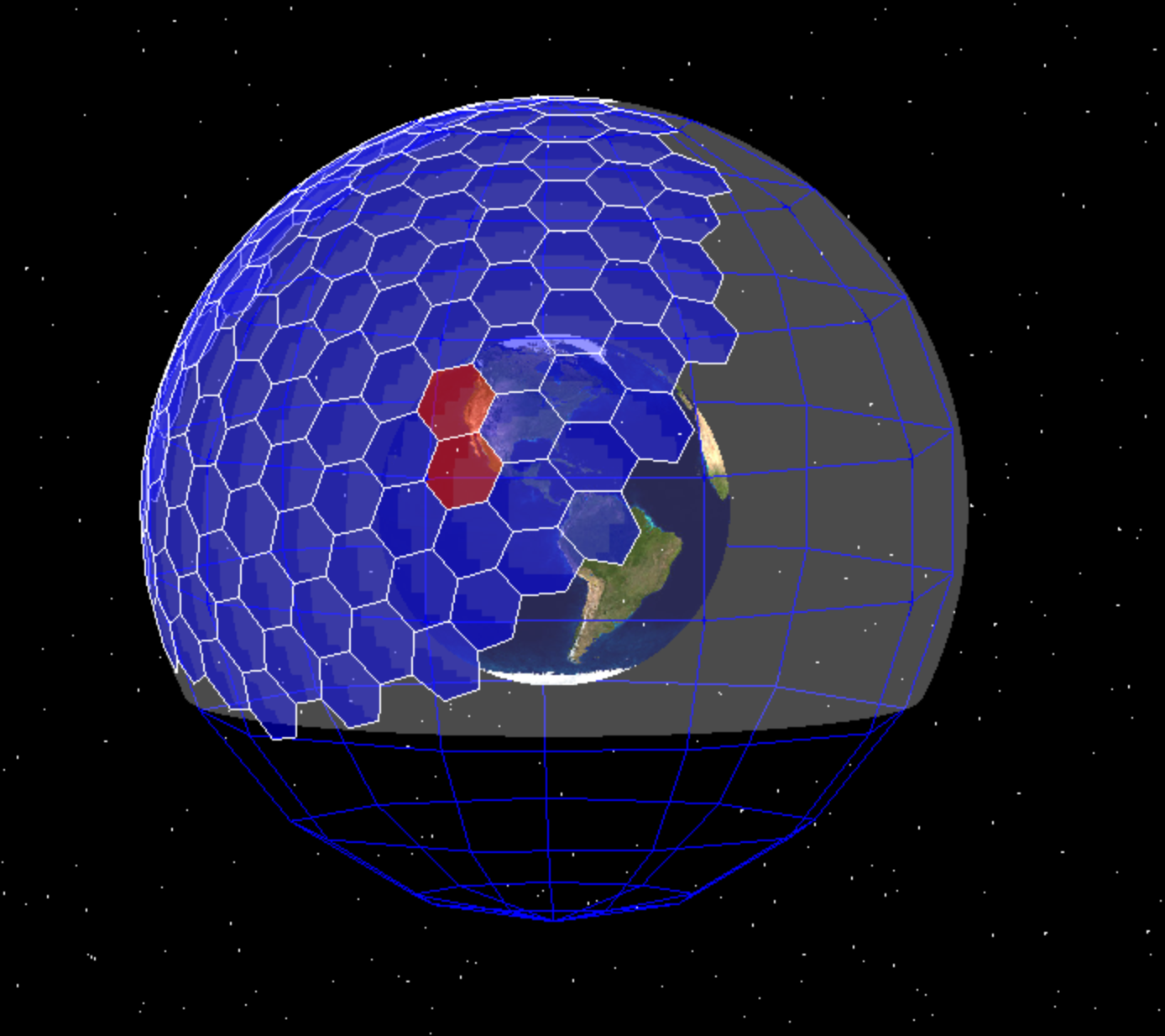}
\includegraphics[width=0.45\textwidth]{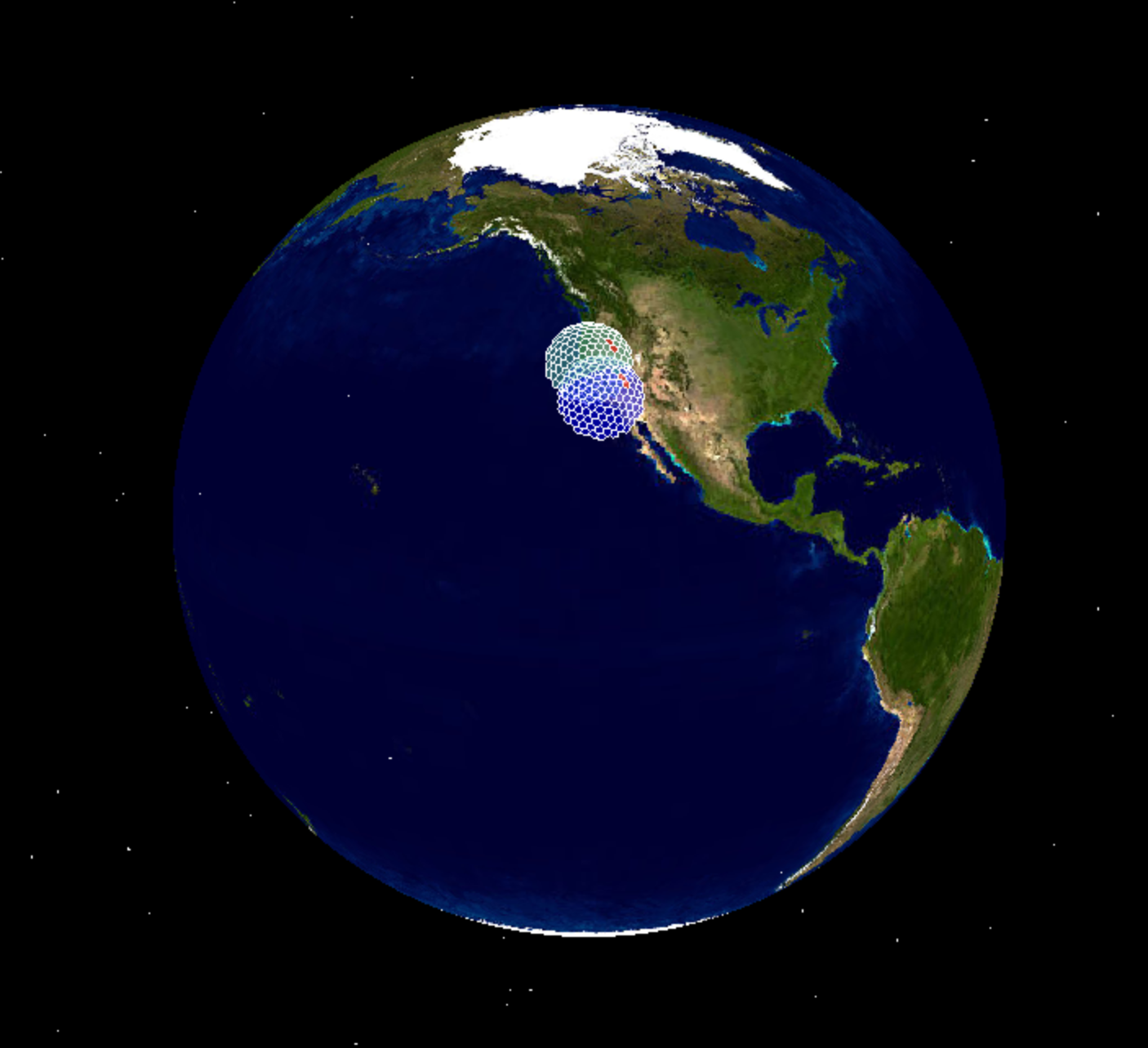}
\includegraphics[width=\textwidth]{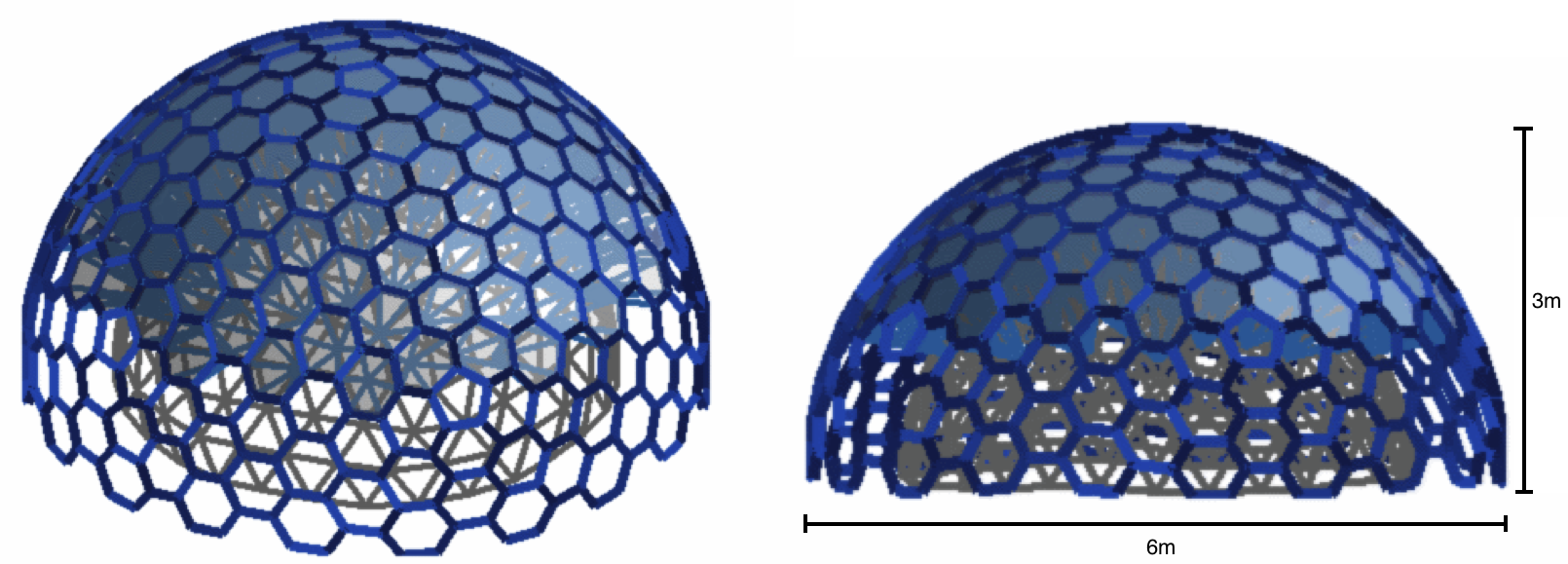}
\caption{Conceptual design of an optical SETI observatory with two geodesic dome assemblies observing the same regions of the sky for coincidence detection, which is essential for confirmation of events. Top-left panel represents the instantaneous FoV of the observatory covering all-observable sky (30$^\circ$ above horizon) in the ideal case of hexagonal detector arrays which could be dedicated to either visible (blue hexagons) or near-infrared observations (red hexagons). Top-right panel shows the two locations identified to cover the Northern hemisphere. A conceptual design of a geodesic telescope assembly is represented on bottom panels where 126 hexagonal 0.5m Fresnel-lens telescopes (shaded hexagons) tile the dome structure.}
\label{fig:concept}
\end{figure}

The frequency of tessellation is the number of segments into which the grid divides the polyhedron edges. Triangular grids can be obtained by running great-circle arcs perpendicular to the edges. If the chosen tessellation frequency is a multiple of 3, triangular grids can then be used to derive hexagonal grids (each hexagon being constituted of 6 merged triangles), with 12 occasional pentagons centered on the icosahedron vertices. As shown in Fig.\ref{fig:concept}, this forms a layout of regular apertures that directly matches the curvature of the sky. Member lengths and joint coordinates define all the pointing directions of the housed telescopes, distributed evenly on a grid as coarse or fine as required by the design. The height and diameter of the dome depend on the number and size of apertures. Since members are straight struts, geodesic domes are economical and well suited for prefabrication and easy assembling. The number of telescopes that a dome can hold is directly related to the tessellation frequency. Table \ref{tab:dome3} gives examples of geometric designs truncated at 30-degrees elevation with different tessellation frequencies giving different instantaneous FoV per telescope. Once the tessellation frequency has been chosen to match the required FoV per telescope, one can scale the dome structure dimensions to match the required aperture area.

\begin{table}[htb]
\caption{Geometric layouts of a 12th, 18th and 24th-frequency geodesic dome. } 
\label{tab:dome3}
\begin{threeparttable}
\begin{center} 
\begin{tabular}{ |l| p{3.3cm}| p{3.3cm}| p{3.3cm}| }
\hline 
Tessellation frequency f& 12& 18 & 24 \\
\hline 
Dome Height [m] & 2.5& 5 & 5 \\
\hline
Number of lenses & 126& 271 & 486 \\
\hline 
Lens diameter\tnote{a} [m] & 0.49 & 0.64 & 0.49 \\
\hline
Min. Field-of-view per lens [$^{\circ}$] &9.2& 6.2 & 4.6 \\
\hline
Total collecting area [m$^2$] & 24.8&72.1 &75.9 \\
\hline
Number of hex members & 414 & 869 & 1,534 \\
\hline
Mean hex members length [m] &0.25 & 0.32 & 0.24 \\
\hline
Number of hex joints & 415 & 835 & 1,425 \\
\hline
Geometrical layout & 
\raisebox{-\totalheight}{ \includegraphics[scale=0.2]{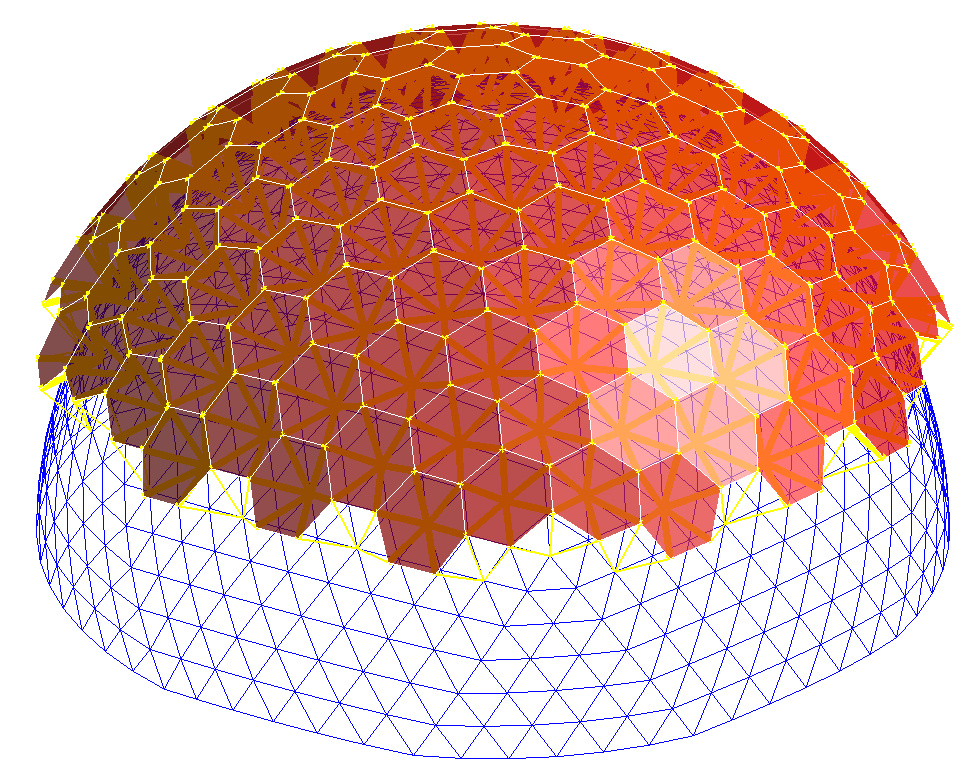}}&
\raisebox{-\totalheight}{ \includegraphics[scale=0.2]{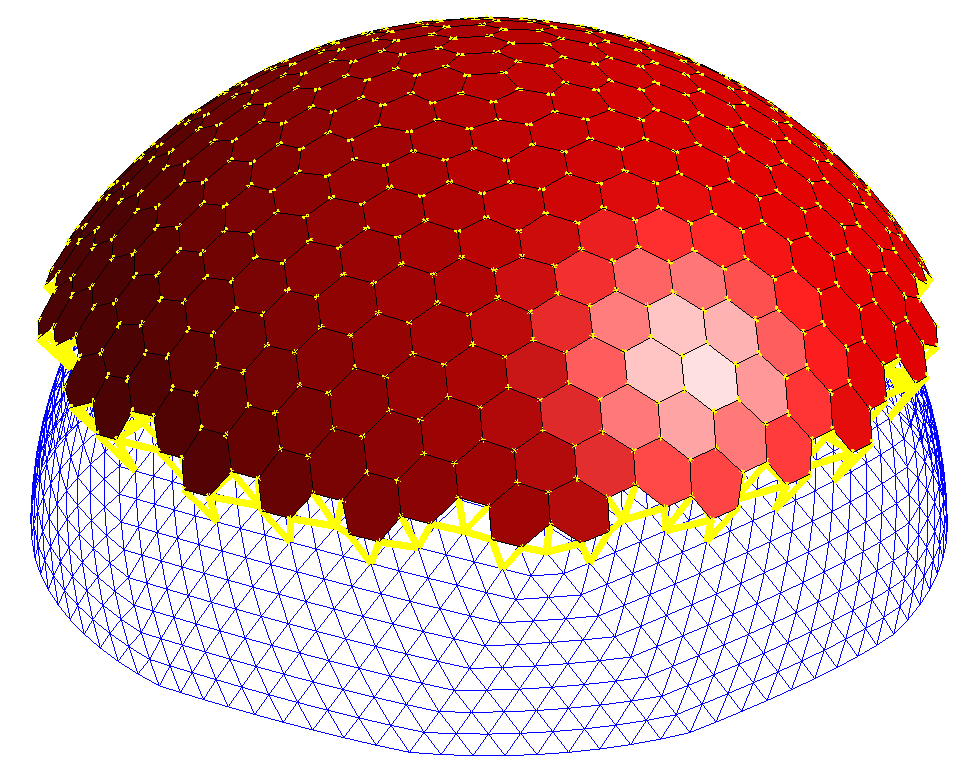}}&
\raisebox{-\totalheight}{ \includegraphics[scale=0.2]{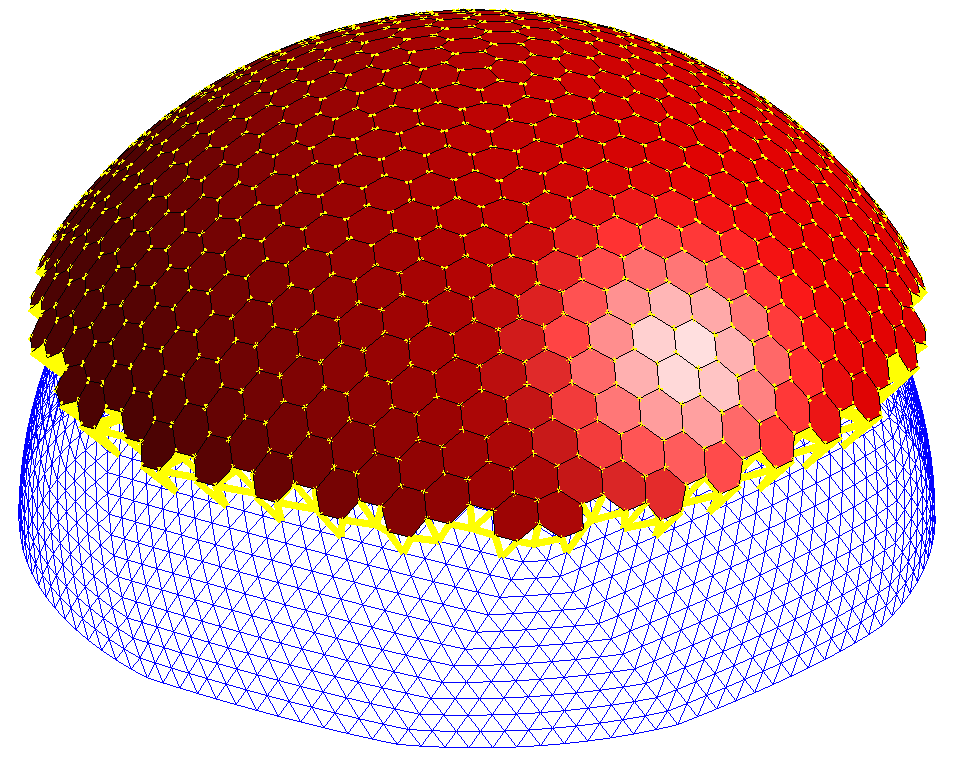}}\\ 
\hline
\end{tabular}
\begin{tablenotes}
\item[a]   calculated from the mean hexagon equivalent area.
\end{tablenotes}
\end{center} 
\end{threeparttable}
\end{table}

\subsection{TESSELLATION COMPARISONS} \label{sec:structural-tessell}
While it is possible to subdivide one flat triangular face of the icosahedron with identical equilateral triangles, the required subdivision should occur on the surface of the spherical icosahedron to maintain the curvature of the structure and the proper orientation of the telescopes. The nodes created by the subdivision define the coordinates of the joints. Straight members connected to these joints are located inside the sphere resulting in different chord lengths. Various subdivision schemes have been developed to form uniform grids of nearly-identical flat triangles\cite{Popko2012}, differing in the choice of great-circle arcs to divide the face. Each subdivision method gives a grid with different chord lengths, triangle areas and shapes. A subdivision method with small variations of these quantities tends to give more uniform grids. 

To subdivide the spherical icosahedron face, we tested three different methods: 
\begin{itemize}
\item  \textit{Equal-chords}, based on the subdivision of the icosahedron edge chord. The equally spaced points formed on the edge chord are located inside the sphere and projected onto the circumscribing sphere, resulting in unequal arc lengths between edge reference points. Great-circle arcs, passing by reference points on two opposite edges of the icosahedron face and almost parallel to the third edge, intersect with other great circle arcs defined with reference points located on an other pair of opposite edges, then subdividing the face into triangles. This method gives the most uniform (near equilateral) triangular faces\cite{Popko2012};
\item  \textit{Equal-arcs (two great circles)}, which differs from the previous method in starting by dividing the spherical icosahedron edges in equal arcs rather than equal chords;   
 \item  \textit{Equal-arcs (three great circles)}, which uses the three pairs of opposite edges to form a set of great circle arcs, which do not meet at the exact same points. Instead, three great-circle arcs meet at three points on the spherical face. A small triangle results and its centroid is projected to the surface of the sphere. This subdivision method gives the smallest variance in member length and triangle area. It also gives the most uniform distribution of face orientations\cite{Popko2012}.  
 \end{itemize}

Several metrics have been defined for comparing one subdivision method with another\cite{Popko2012}. Since our final subdivision could be made of hexagons (each one formed with 6 merged triangles), we used the following metrics for comparing subdivision methods of the icosahedron for a given tessellation frequency:
\begin{itemize}
\item number of unique chords, which should be as small as possible for fast manufacturing of the members. For each subdivision method, member lengths were sorted with 1-mm bin width and the resulting number of unique chords is reported in Table \ref{tab:metrics};
\item uniformity quotient, or the ratio of the longest to the smallest hex member lengths; 
\item standard-deviation of chord lengths, which should be as small as possible to form the most uniform grid; 
 \item coplanarity of the points forming the hexagons. The hexagons derived from the triangular grid, unless special measures are taken, are not perfectly planar. The installation of modules inside the hexagons requires them to be as planar as possible in order to facilitate the construction and assembly of the supporting edge frame. For each subdivision method, we calculated deviations of the vertices of each hexagon from their average plane to determine the hexagon flatness. Mean and standard deviation of the distance from average hex plane are also reported in Table \ref{tab:metrics};
 \item largest inscribing circle that could fit inside any hexagons of the grid (pentagon excluded). The larger the diameter, the more evenly distributed the hex points are.
\end{itemize}

 \begin{table}[ht]
\begin{center}
\begin{tabular}{|l|l|l|l|l|l|l|} 
\hline
\rule[-1ex]{0pt}{3.5ex} & \parbox[b]{1.2cm}{Unique\\chords$^a$} & \parbox[b]{1.6cm}{Uniformity\\quotient} & \parbox[b]{1.6cm}{Uniformity\\STD [cm]} & \parbox[b]{2.2cm}{Coplanarity\\ distance [mm]} & \parbox[b]{1.8cm}{Coplanarity\\ STD [mm]} & \parbox[b]{2.6cm}{Inscribing circle\\   diameter [m]} \\
\hline
\rule[-1ex]{0pt}{3.5ex}  \textit{Equal-chords} & 11 &  1.293  & 2.01 & 0.54 & 0.38& 0.424\\
\hline
\rule[-1ex]{0pt}{3.5ex} \parbox[b]{3cm}{\textit{Equal-arcs with}\\\textit{two great circles}}  & 32  & 1.217  & 1.67& 1.22 &   0.82&0.452\\
\hline
\rule[-1ex]{0pt}{3.5ex} \parbox[b]{3cm}{\textit{Equal-arcs with}\\\textit{three great circles}} & 14 & 1.1743  & 1.14 & 1.17 &0.74&0.462 \\
\hline
\end{tabular}
\end{center}
\vspace{-0.25cm}\hspace{0.5cm}  a: 1-mm bin width 
\caption{Comparisons of different subdivision methods for a 3m-high 12th-frequency hexagonal tessellation. } 
\label{tab:metrics}
\end{table} 

Our simulations of 3-m, 12-th frequency hex dome show that the \textit{Equal-arcs (three great circles)} method gives a more uniform grid, resulting in the largest inscribed circle diameter (Table \ref{tab:metrics}). The \textit{Equal-chords} method gives the most flattened hexagons and least number of unique chords. \textit{Equal-chords} and \textit{Equal-arcs (three great circles)} coplanarities are within acceptable values for the installation of modules on a 3-m high dome structure.

\subsection{STRUCTURAL ANALYSIS} \label{sec:structural-analysis}
\begin{figure}[ht] 
\includegraphics[width=0.7\textwidth]{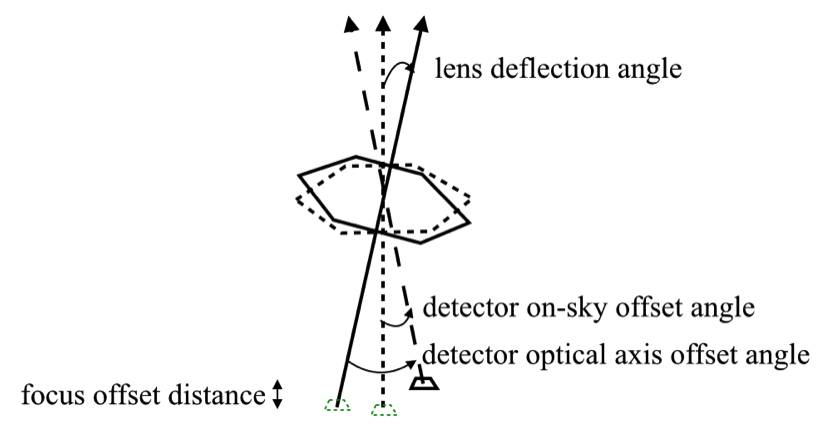}
\caption{ Displacements of components (hexagonal lens and square detectors), due to external forces, can be related to optical parameters such as the lens deflection angle, measuring the angle made by the modified pointing direction of the lens with its expected pointing direction when no external forces are applied (small-dashed line). Structural analysis could also determine the focus offset distance and detector on-sky offset angle due to the detector displacement. }
\label{fig:opticalstruc}
\end{figure}

Dome structures are impressively strong with an even distribution of weight that efficiently distributes stress along their structures. Domes can mechanically withstand extreme wind, but PANOSETI will be further protected from bad weather by a strong enclosure. Geodesic domes use much less material than conventional buildings, yet outperform them in structural tests. The dome design, material, thickness and shape of members and joints can be modified to increase the rigidity of the structure. Those parameters constrain the maximal displacement of joints and members when external forces are applied. We identified optical parameters affected by joint displacements. These parameters, like the lens deflection angle, are represented on Fig.\ref{fig:opticalstruc} showing the geometry of lens and detector, deformed by external forces.
 
We implement a finite element method (the so-called stiffness method) to calculate displacements of dome elements (members and joints) under wind and gravity loads, as well as thermal expansion, to determine optimal joint design as well as member material and thickness. These displacements can be constrained to have a minimal impact on optical performance of the telescope modules.

Simple geodesic domes are usually based on a triangular tessellation,
ensuring their rigidity and stability. Large geodesic domes made of regular hexagons may need additional sources of structural support, such as an inner triangular layer connected to the hexagon layer with in-between struts or tension rods. Together, the outer and inner layers could act as a double-layer spherical truss system (e.g. the 61-m high Biosphere of Montreal, Canada or The Eden project in Cornwall, England). The reasonably open nature of the structure between the two layers allows installation of telescope modules within the structural depth. Their mounting is simplified as there is a regular system of supports available. Our on-going structural analysis should determine whether a dual-layer structure is required. It will also help to understand the tolerances required when building a dome, and experiment with ways of improving the structural soundness of the telescope assembly.

\section{OPTICAL DESIGN}\label{sec:optical}
Single-aperture wide-field astronomical instruments usually have complex optical designs aimed at minimizing optical aberrations and maintaining image quality over large FoV\cite{Angel2001}. This can be obtained with wide-field correctors which often require large, heavy and costly optical components\cite{Ackermann2010}. Sensitivities of optical pulsed SETI experiments depend on the size of the collecting apertures and while it would be ultimately important to know with high accuracy the celestial coordinates of a positive detection, optical SETI initial searches could be performed at low-angular resolution, relaxing the constraints on the required optical quality. Once a positive detection has been noticed and repeated, more precise localization could be performed with high-angular resolution instruments like OSETI\cite{Stone2005, Howard2007} and NIROSETI\cite{Wright2014}. Low-angular resolution ($\sim$ few arcminutes) large-collecting apertures ($>$0.4m), such as refractive Fresnel lenses, have been used for the detection of nanosecond optical pulses generated by high-energy cosmic-rays and gamma-rays striking the top of the Earth's atmosphere \cite{Bunner1967,Bunner1968,Stephan2013,Fujii2018}.
 
We propose to use large refractive Fresnel lenses in conjunction with high-speed detectors for performing low-angular resolution optical SETI, allowing for a cost-effective, lightweight alternative to conventional lenses and mirrors. Since the refractive power of a lens is set only by its optical interfaces, one can divide the continuous surface of a standard lens into a set of surfaces of the same curvature (e.g. concentric grooves), with step-wise discontinuities between them, allowing the removal of lens material between the interfaces while maintaining the lens surface curvature. Considerably lighter than conventional lenses of the same size, these Fresnel lenses are ideally suited for a large-scale assembly structure. Refractive Fresnel lenses can be designed with small f-numbers (typically between 0.5 and 2), which make them well-adapted for observing large FoV at low-angular resolution, using large detector pixel sizes such as millimetric Multi-Pixel Photon Counter (MPPC) and Discrete Amplification Photon (DAPD) detectors selected for the PANOSETI experiment\cite{Wright2018}. Refractive Fresnel lenses designed for visible and near-infrared applications are often made of optical-grade acrylic (PPMA) which have high transmittance ($>92\%$ between 400-1700\,nm)\cite{Davis2011b}, good stiffness and abrasion resistance, high resistance to UV radiation, and low water absorption in comparison to other polymers. Different techniques of fabrication are used such as injection or compression molding, or hybrid methods (such as coining), with different levels of precision and replication fidelity\cite{Davis2012}.
 
\begin{figure}[htb]
\begin{subfigure}{0.45\textwidth}
\centering
 \includegraphics[width= 0.85\textwidth]{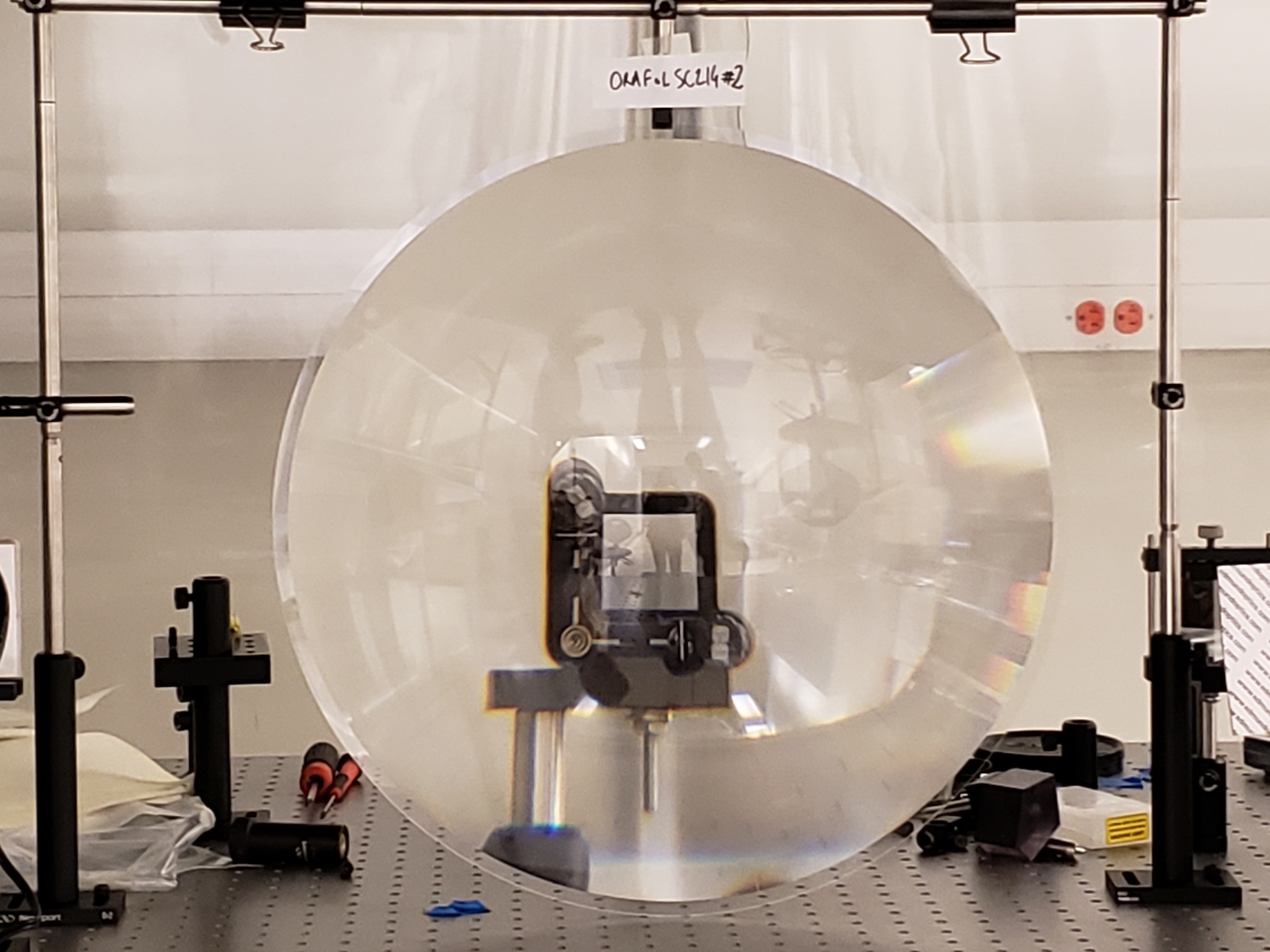}
 \end{subfigure}
 \begin{subfigure}{0.45\textwidth}
\includegraphics[width= 1.05\textwidth]{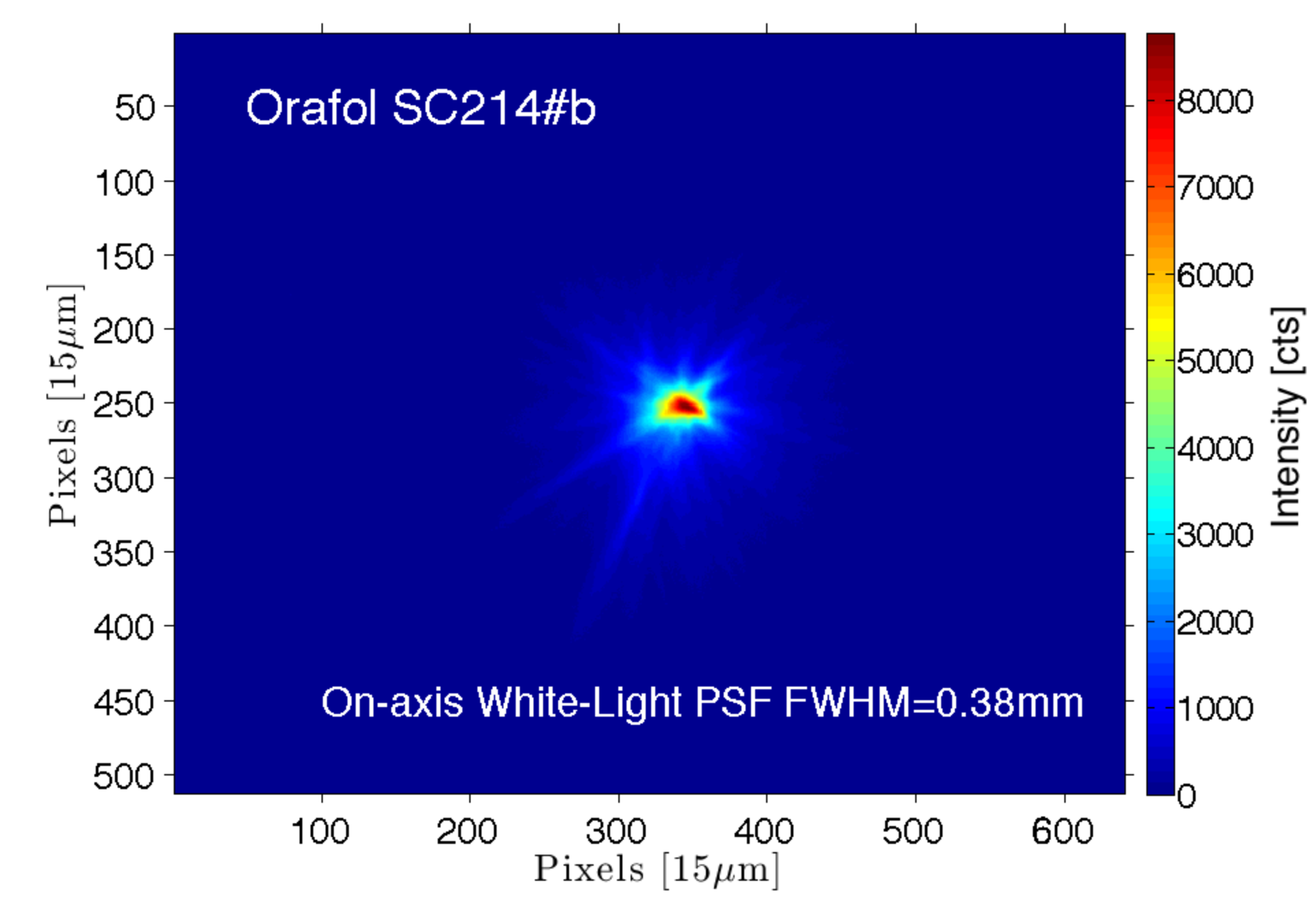}
\end{subfigure}
 \begin{subfigure}{0.45\textwidth}
 \includegraphics[width=1.04\textwidth]{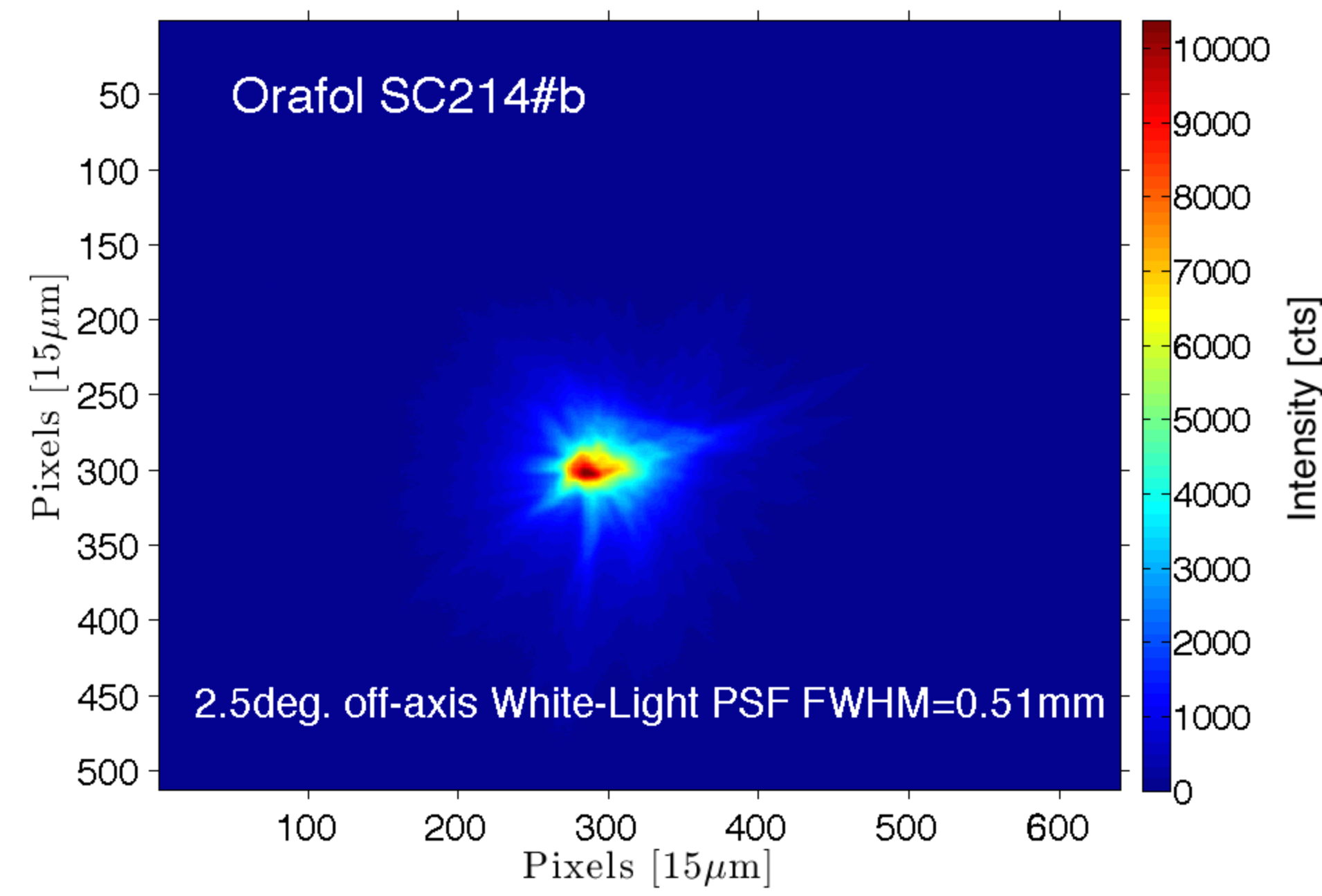}
 \end{subfigure}
 \begin{subfigure}{0.45\textwidth}
\includegraphics[width= 1.05\textwidth]{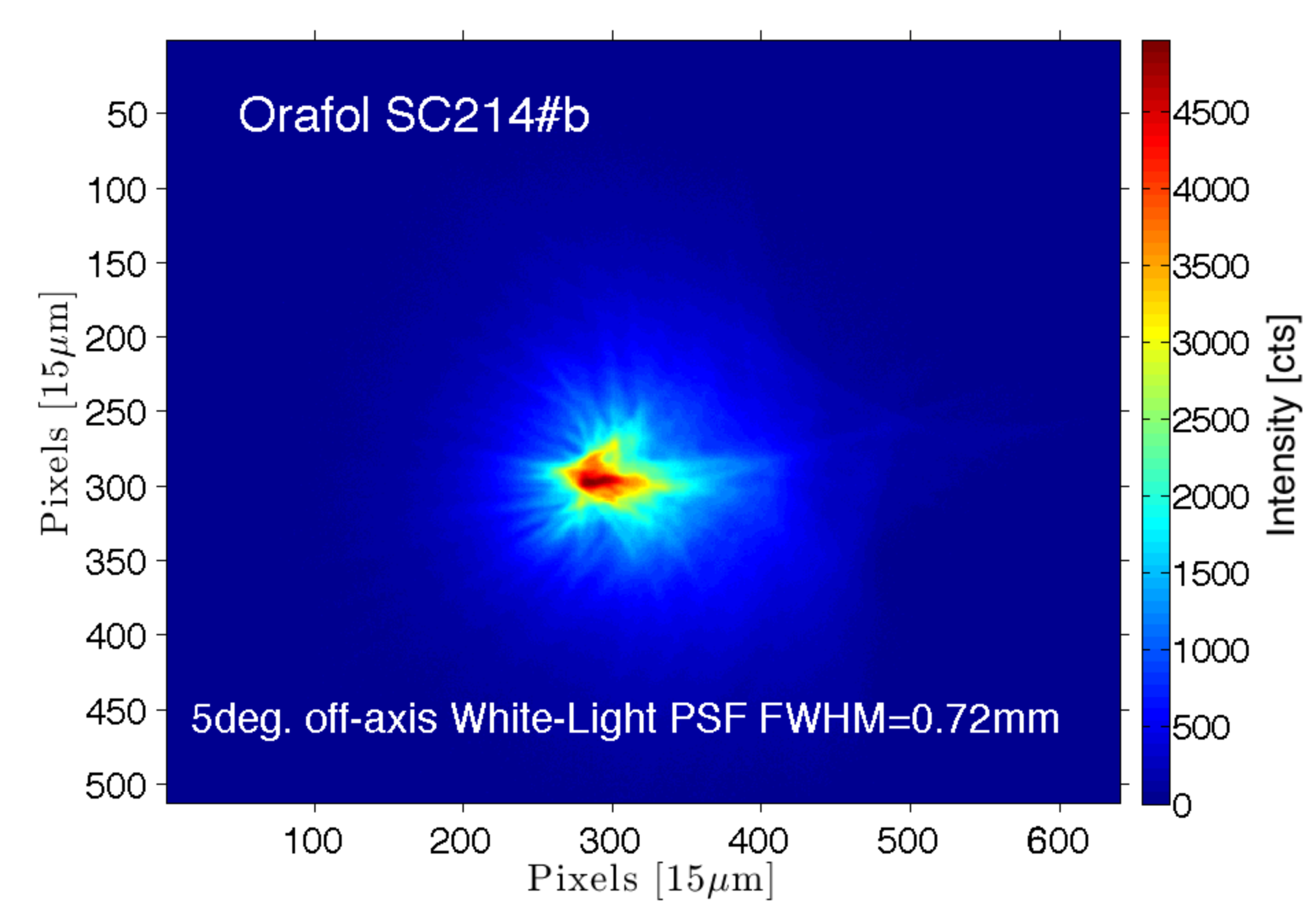}
\end{subfigure}
\caption{Top-left: Refractive Fresnel lens Orafol SC214 mounted on the optical table for spot size measurements. Top-right: On-axis spot obtained by imaging a white-light point-source (white LED with a 1mm pinhole located 7m from the lens). The green color represents pixels containing half of the maximal intensity. Bottom-left: white-light spot from $2.5^\circ$ off-axis point-source. Bottom-right: white-light spot from $5^\circ$ off-axis point-source.}
\label{SC214}
\end{figure}

\begin{figure}[htb]
\begin{subfigure}{0.45\textwidth}
 \includegraphics[width= \textwidth]{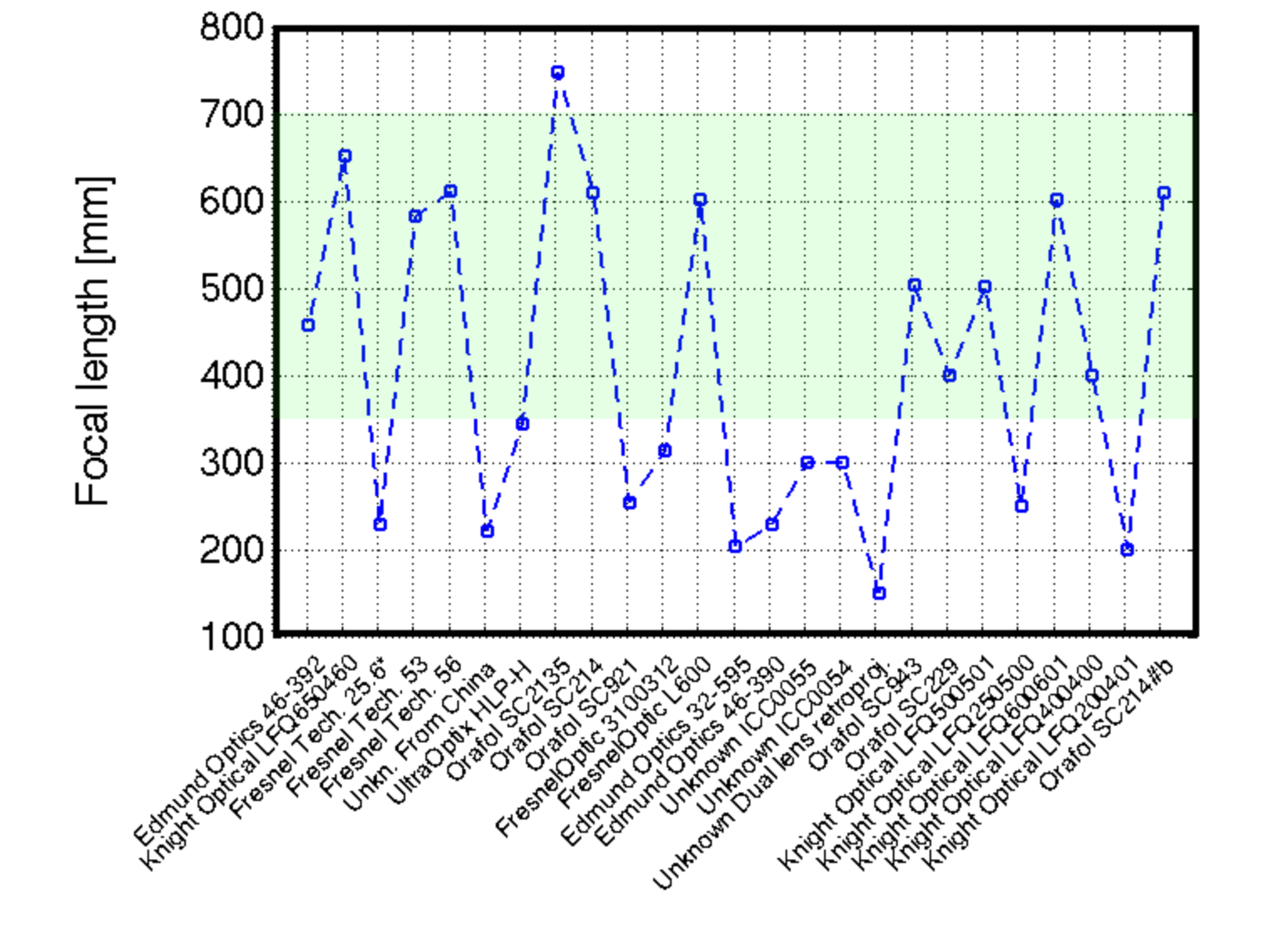}
 \end{subfigure}
 \begin{subfigure}{0.45\textwidth}
\includegraphics[width= \textwidth]{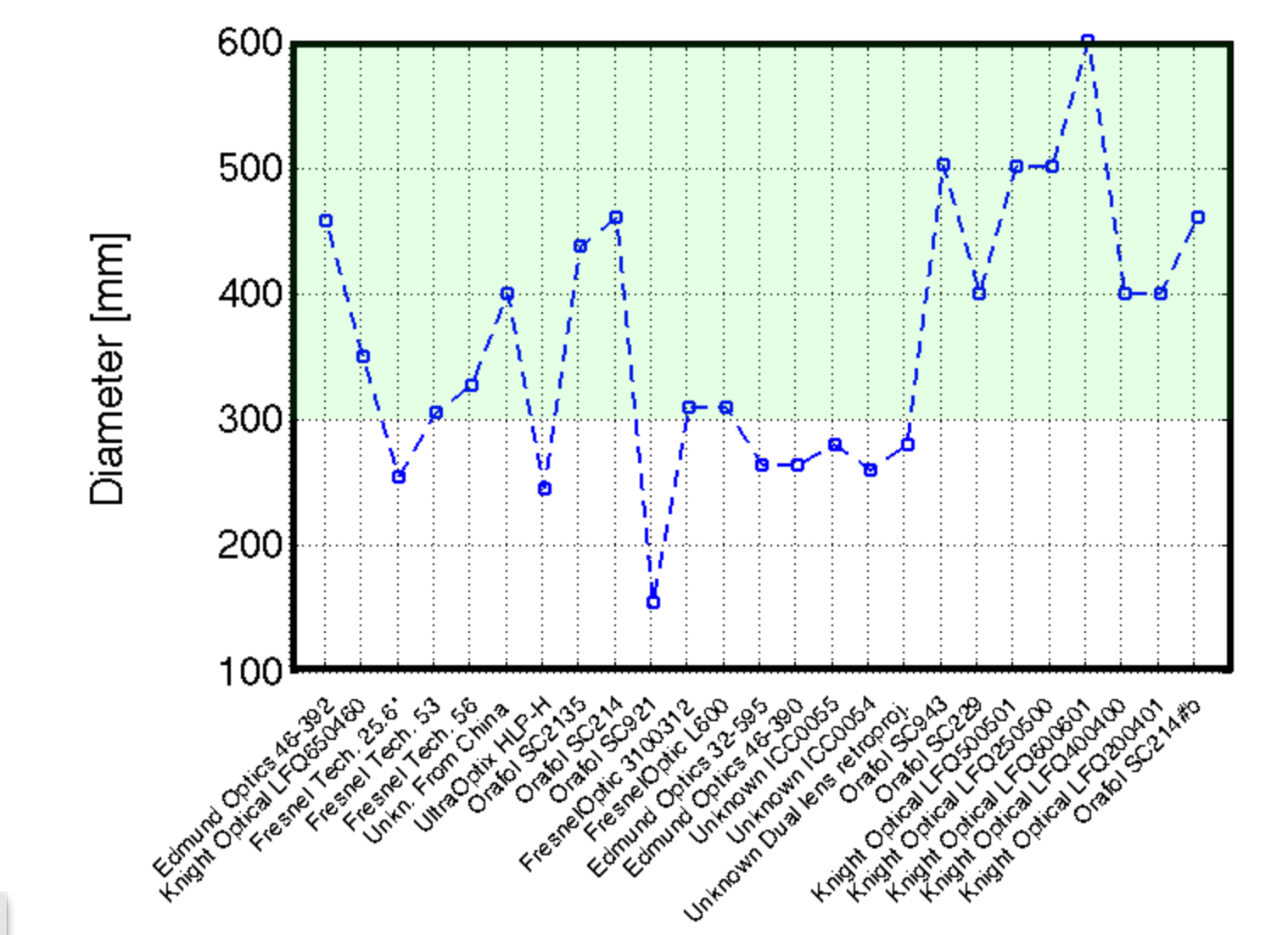}
\end{subfigure}
 \begin{subfigure}{0.45\textwidth}
 \includegraphics[width= \textwidth]{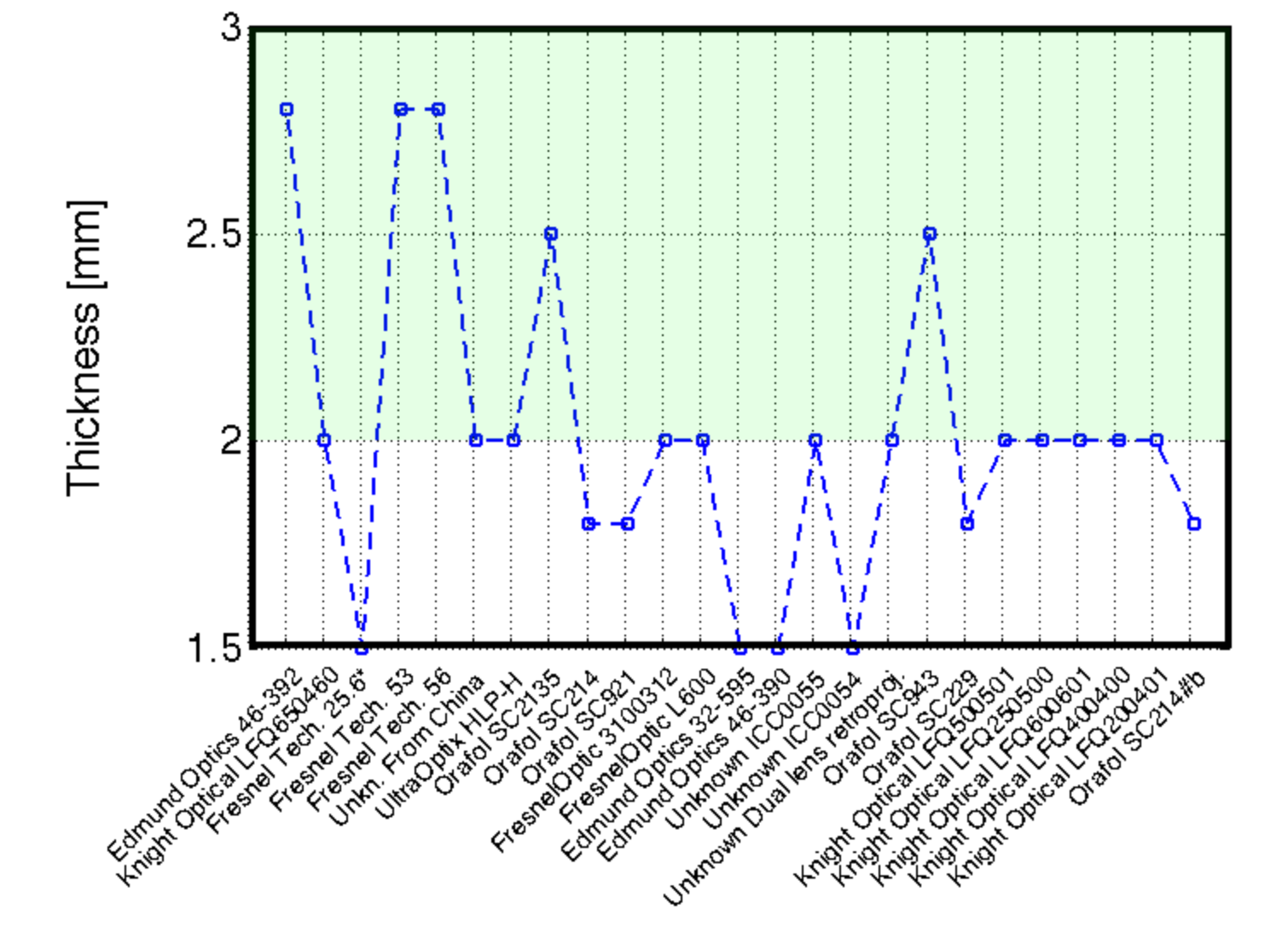}
 \end{subfigure}
 \begin{subfigure}{0.45\textwidth}
\includegraphics[width= \textwidth]{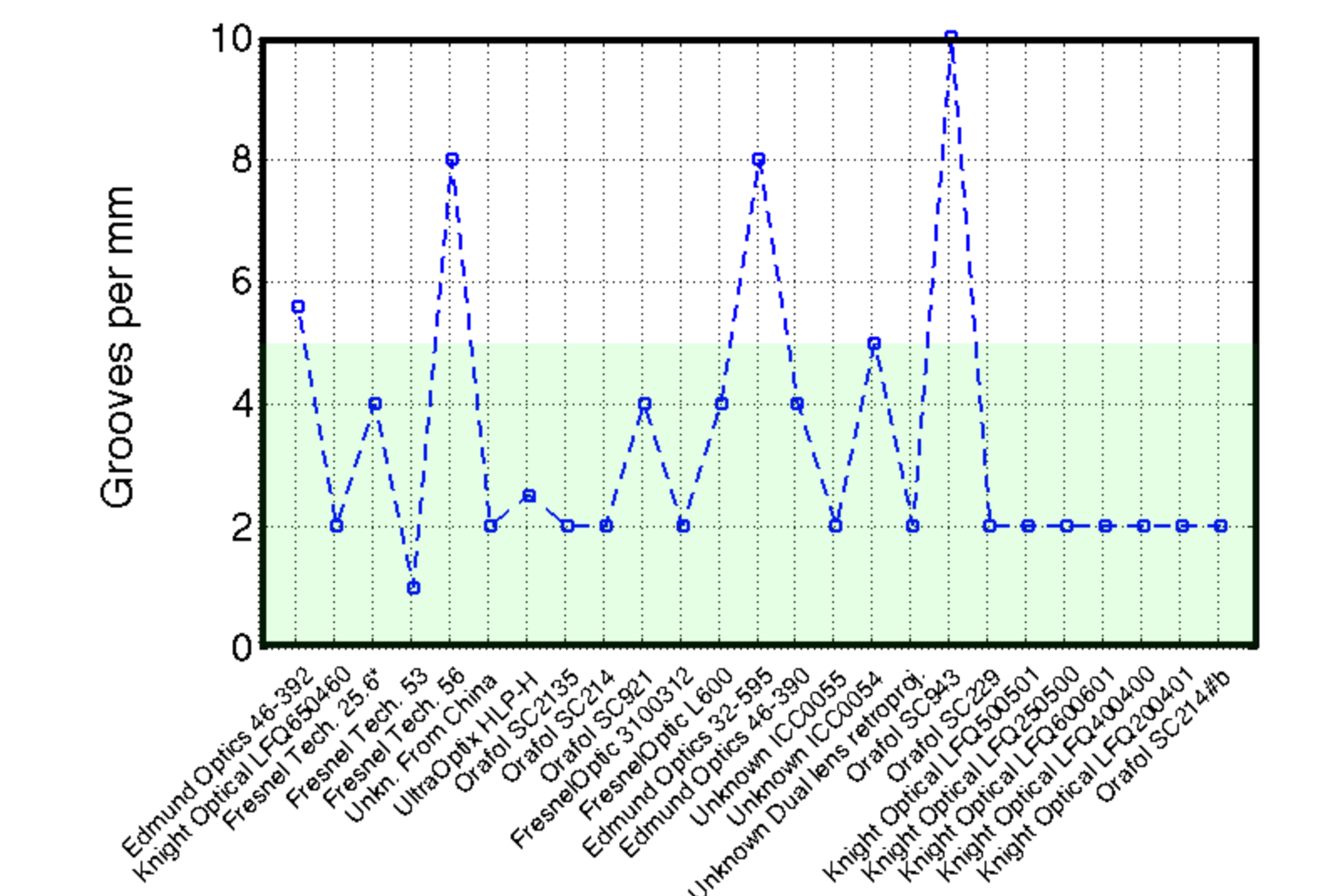}
\end{subfigure}
\begin{subfigure}{0.45\textwidth}
 \includegraphics[width= 1.05\textwidth]{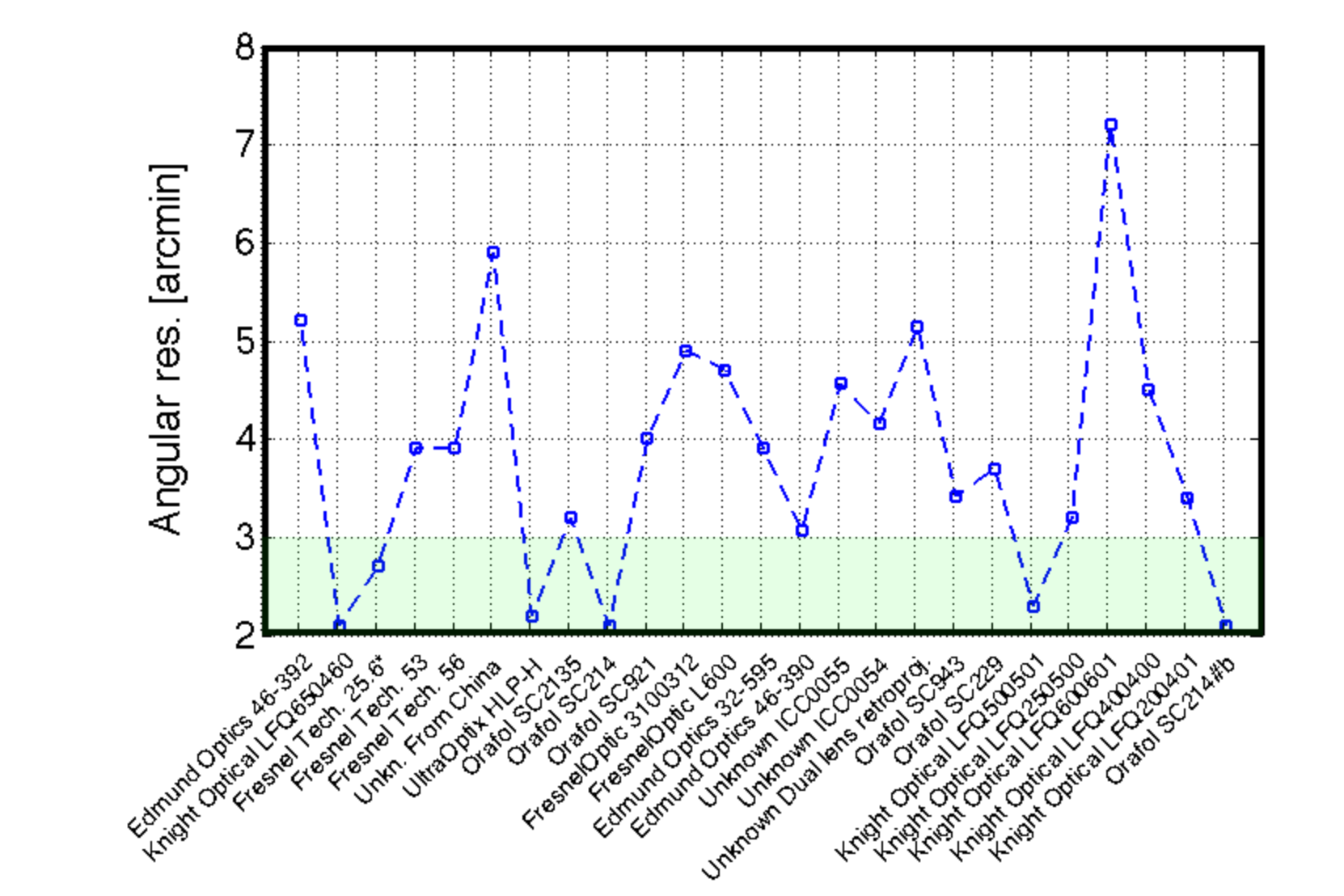}
 \end{subfigure}
 \begin{subfigure}{0.45\textwidth}
\includegraphics[width= \textwidth]{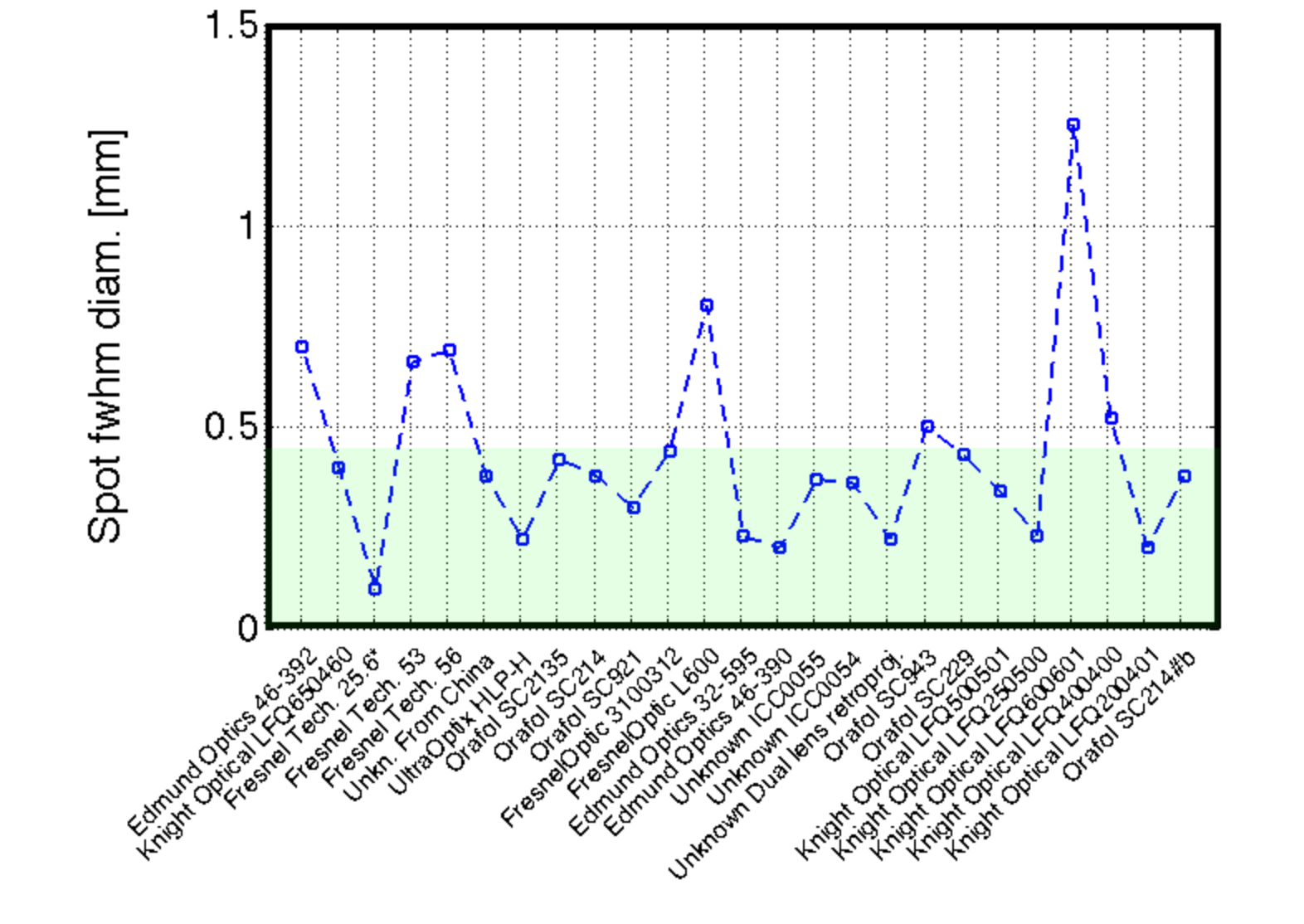}
\end{subfigure}
\caption{Characterization of off-the-shelf refractive Fresnel lenses in terms of their focal lengths (at 0.55$\mu$m), diameter of their grooved surfaces, thicknesses and groove widths. Angular resolutions and spot sizes were obtained by imaging an on-axis white-light point-source. }
\label{allres}
\end{figure}

Although modern materials, new molding techniques, and computer-controlled diamond turning machines have improved the optical quality of Fresnel lenses, allowing them to be built with complex aspheric surfaces, refractive Fresnel lenses have inherent optical aberrations that need to be taken into account and corrected as part of design or selection considerations. Various types of aberrations can affect the on-axis point-source spot size and intensity.
For instance, the throughput of the lens is inherently affected by geometric losses due to internal reflections at the groove step discontinuities, with this effect being more pronounced with increasing distance from the lens center. For ease of manufacturing the facets of the groove are often made flat, but newer methods can be used to cut each groove facet in the correct curved contour, thus avoiding even the width of the groove (typically 0.1 to 1 mm, see Fig.\ref{allres}) as a limit to the sharpness of focus. Another way to reduce the effect of facet flatness is to reduce the groove width. However, as the pitch becomes smaller, the facets will become more efficient at acting like a grating sending light into higher diffractive orders away from the desired focal position. Other manufacturing defaults, such as rounding of the groove tip \cite{Davis2011}, can contribute to the loss of sharpness and lens transmittance. Bowing of the lens, due to its own weight, change of temperature, or external forces such as wind load can affect the optical quality of the lens. If the lens is supported at its edges, the maximal deflection occurs at the center of the lens, modifying the location of the image plane along the optical axis. Increasing the thickness of the lens or adding stiffener beams on its surface can significantly reduce deflections\cite{Cosens2018}.
 
Chromatic aberrations affect all single-element lenses, and conventional refractive Fresnel lenses are not an exception. Because of the similar refractive index and dispersion values of acrylic and common crown glasses (like BK-7), acrylic Fresnel lenses suffer from similar chromatic aberrations. Location of the focal plane along the optical axis varies with wavelength, so the detector array receives well focused rays at a specific wavelength, and a section of a light cone (for circular lens) at different wavelengths. PANOSETI detectors (or ``pixels"), acting like ``light buckets", can receive all photons at defocused wavelengths if those pixels are larger than the cone section. Although chromatic aberrations limit the effective wavelength range of the instrument, PANOSETI pixels are large enough to cover 300nm of visible bandwidth ($>$ 50\% transmission) when used with a 0.5-m f/1 acrylic lens, as well as near-infrared Y, J, and H-bands, if observed separately. To expand the wavelength coverage, achromatization of the system is under investigation. Correction of chromatic aberration is usually performed by adding a lens made of material with a different Abbe number and opposite power. The lenses can be in contact\cite{Vallerotto2016} as in a conventional achromatic doublet or separated \cite{Faklis1989,Early2004}, but usually require large correctors especially for low f-number applications. Achromatization could also be achieved by use of hybrid Fresnel lenses or by segmenting the focal plane into manageable fields where aberrations could be corrected by using smaller lenses and different pupil plane corrector plates\cite{Ragazzoni2004}.

We compared on- and off-axis optical performance of several off-the-shelf refractive Fresnel lenses with low f-number ranging from 0.5 to 2. Lenses were supported vertically by four points of support to prevent bowing. We used a VIS-NIR InGaAs 14-bit camera (with 15$\mu$m pixel pitch) to capture the image of a 1-mm pin-hole object located at 7m from the lens (30-arcsec diameter) illuminated with visible LEDs (white, red, green) or CW laser-diodes working at 0.63$\mu$m, 1.31$\mu$m, and 1.55$\mu$m. Exposure times were set so as not to saturate any pixel, and the camera was translated along the optical axis for focus adjustment. The light source system was moved for off-axis characterizations at 2.5 and 5 degrees off-axis. We recorded the number of pixels having intensities higher than half the maximal intensity to calculate the diameter of an equivalent circular area. Images obtained with the Orafol SC214 lens and the white light source are represented on Fig.\ref{SC214}, showing 0.3-mm FWHM ($\sim$ 2.1 arcmin) spot size for on-axis observations, increasing to 0.5mm ($\sim$ 3 arcmin) and 0.7 mm ($\sim$ 4.1 arcmin) for, respectively, 2.5-deg and 5-deg off-axis measurements. Fig.\ref{allres} shows the design parameters of several commercial lenses tested in the lab along with measured angular resolutions and spot sizes. Their focal lengths, which set the plate scale of our instrument, should be small enough to reduce the number of detectors needed for covering a given instantaneous FoV, but the ratio of the f-number of the system to the pixel area should be large enough to limit the decrease in sensitivity due to sky background photons. Orafol SC214 and Knight Optical LFQ650460 Fresnel lenses meet our requirements, with the finest angular resolution and largest aperture diameter.

\section{FIELD-OF-VIEW CONSIDERATIONS}\label{sec:fov}
In order to maximize sky coverage of the geodesic assembly of telescopes, it is necessary to optimize on-sky position angle of the detectors. To avoid redundant pixels or gaps between adjacent telescope FoVs, detector arrays would need to have the same shape as the regular polygon used to tile the geodesic dome structure (e.g. hexagon in the design considered in this paper). However, for ease of manufacturing, PANOSETI will have square detector arrays\cite{Wright2018} in all identical telescopes. Each telescope will be fixed to the dome and their detector arrays rotated to the proper position angle to maximize sky coverage.

Several mapping parameters were considered to search for an optimal arrangement of square detector arrays. Due to the rotation of each detector array around its telescope optical axis (detector position angle), there is an unlimited number of possible on-sky mapping arrangements. We found it optimal to align detector arrays in such a way that their position angles -- defined as the angular offset of the detector FoV side to the north celestial pole -- will be null for all telescopes, allowing a convenient alignment for star trail follow-up. On-sky coverage is also dependent on the plate scale, the number of detectors (or ``pixels") per array, observatory locations, and tiling arrangements of the apertures (for instance the tessellation frequency changing the number of apertures) as shown in Table \ref{tab:onskymapping}.
 
To calculate redundant areas and effective FoV of the entire telescope assembly, for each design we first calculated the celestial coordinates of all telescope pointing directions determined from the normals of the hexagons (average plane) supporting the telescopes. Plate scale and detector size are then used to calculate the celestial coordinates of the detector array corners. A special algorithm was developed to calculate redundant sky areas, which consists of using a fine grid of binary null cells (each cell representing 1/20th-by-1/20th degree) covering the whole sky, filled with 1's when cells are found to be inside the FoV of a detector array. This gives a binary map that provides the total instantaneous coverage of the assembly and redundant coverage when integrated.

\begin{figure}[htb]
\includegraphics[width=\textwidth,left]{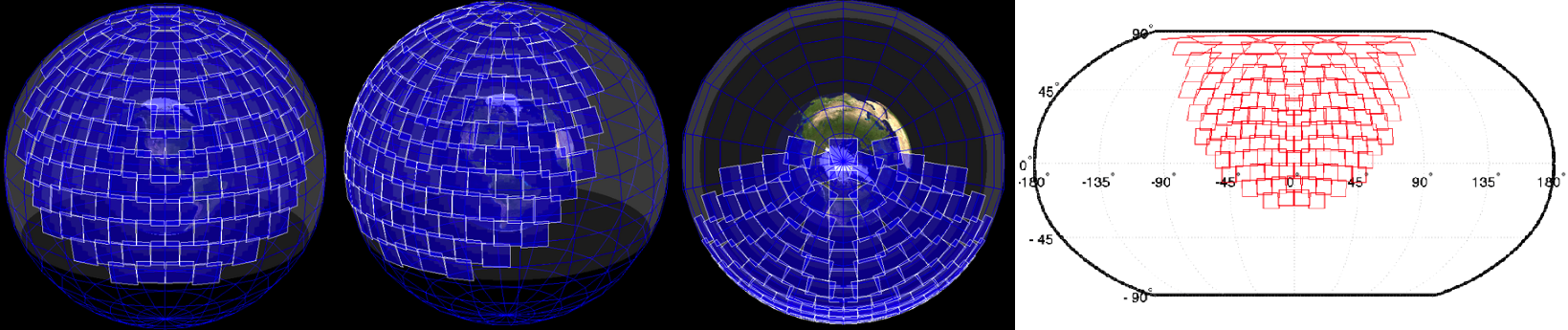}
\includegraphics[width=0.975\textwidth,left]{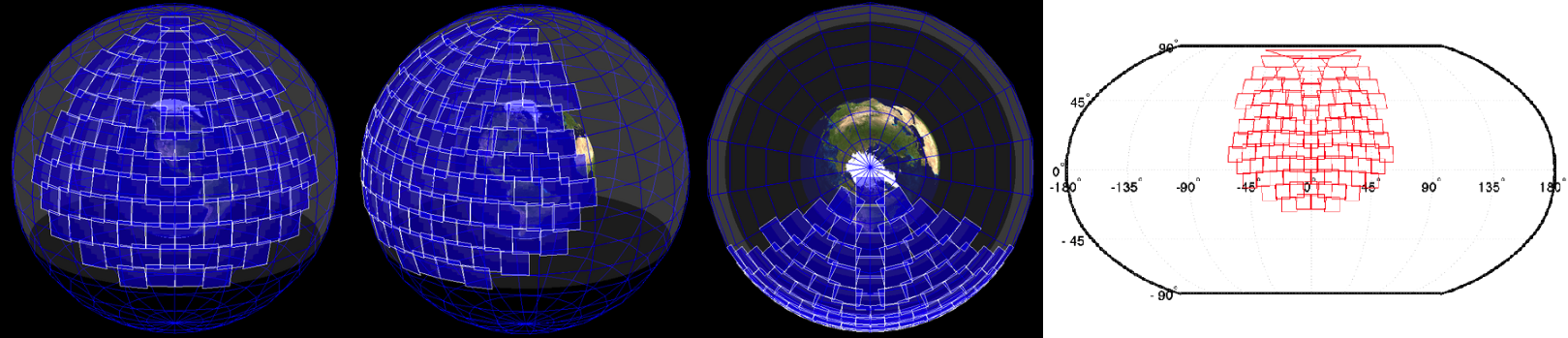}
\includegraphics[width=0.95\textwidth,left]{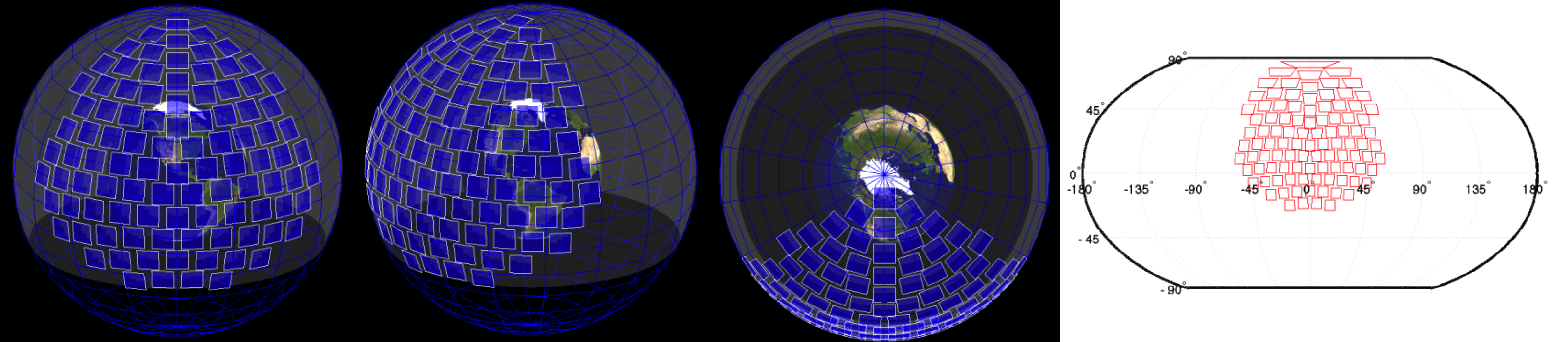}
\includegraphics[width=0.95\textwidth,left]{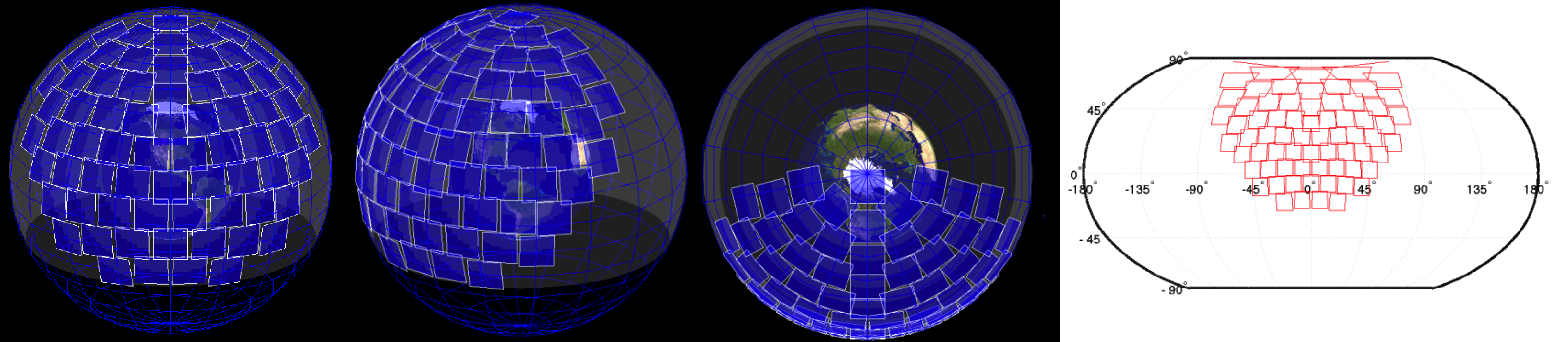}
\includegraphics[width=0.965\textwidth,left]{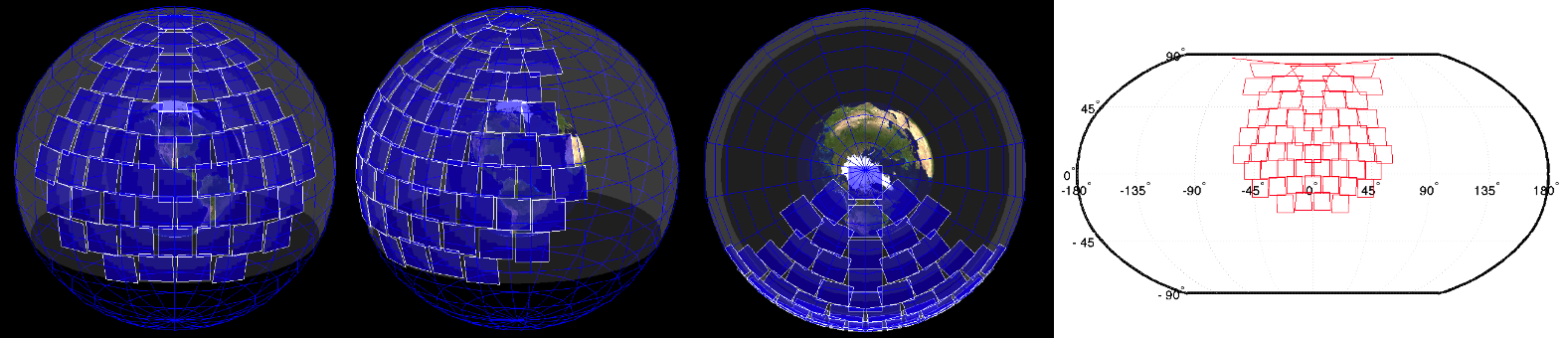}
\caption{Effective sky coverage for different configurations of telescope assemblies. Each row, from top to bottom, illustrates the sky seen for each configuration from configuration A to E respectively.} 
\label{fig:onskymapping}
\end{figure}

A few examples reported in Table \ref{tab:onskymapping} were considered to help specify the structural and optical PANOSETI design parameters. Configurations A and B, which only differ by the number of telescopes in the assembly, give instantaneous FoVs of 10,701 square degrees (25.9\% of the entire sky) and 8,445 square degrees (20.4\% of the sky) respectively (top two rows of Fig.\ref{fig:onskymapping}). Instantaneous FoVs between adjacent telescopes can overlap allowing $\sim$ 15\% of the detectors (``pixels") to be used for relative alignment. Configuration C (middle row of Fig.\ref{fig:onskymapping}) aimed to reduce the number of required pixels by a factor of four by using arrays of 16x16 3mm detectors and a larger plate scale, resulting in a total instantaneous FoV of 5,382 sq. degrees with no redundancy. Configurations D and E (last two rows of Fig.\ref{fig:onskymapping}) reduce the number of apertures by using a 9th frequency tessellation compensated by larger FoV (12 degrees per telescope), a configuration which is more constraining in terms of required off-axis optical quality.
 
\begin{table}[htb]
\caption{Effective sky coverage and redundancy for different configurations of telescope assemblies. } 
\label{tab:onskymapping}
\begin{center} 
\begin{tabular}{|l|l|l|l|l|l|}
\hline
\rule[-1ex]{0pt}{3.5ex} Configuration & A & B &C&D &E \\
\hline
\rule[-1ex]{0pt}{3.5ex} No. of pixel per telescope & 32x32 & 32x32 & 16x16 & 32x32 & 32x32\\
\hline
\rule[-1ex]{0pt}{3.5ex} No. of telescopes & 126 & 99 & 99& 71 & 57\\
\hline
\rule[-1ex]{0pt}{3.5ex} Tessellation frequency$^a$ & 12 & 12 & 12 & 9 &9\\
\hline
\rule[-1ex]{0pt}{3.5ex} Total No. of pixels & 129,024 & 101,376 & 25,344 &72,704& 58,368\\
\hline
\rule[-1ex]{0pt}{3.5ex} Pixel size$^b$ [mm] & 3.28 & 3.28 & 3.2 & 3.28 & 3.28\\
\hline
\rule[-1ex]{0pt}{3.5ex} Focal length$^c$ [m] & 0.6 &0.6 & 0.4 & 0.5& 0.5\\
\hline
\rule[-1ex]{0pt}{3.5ex} Single-aperture FoV & 9.9x9.9deg & 9.9x9.9deg & 7.3x7.3deg & 12x12deg & 12x12deg\\
\hline
\rule[-1ex]{0pt}{3.5ex} Plate scale [deg. per pix.] & 0.31 & 0.31 & 0.458 & 0.376 & 0.376 \\
\hline
\hline
\rule[-1ex]{0pt}{3.5ex} PANOSETI & 10,701 sq.deg. & 8,445 sq.deg. & 5,382 sq.deg. & 9,482 sq.deg. & 7,824 sq.deg.\\
\rule[-1ex]{0pt}{3.5ex} instantaneous & 3.2597 sr & 2.5726 sr & 1.6395 sr & 2.9766 sr & 2.3835 sr \\
 \rule[-1ex]{0pt}{3.5ex} field-of-view & 25.94\% of the sky & 20.47\% & 13.05\% & 23.69\% & 18.97 \% \\
\hline 
\hline 
\rule[-1ex]{0pt}{3.5ex} Redundant coverage & 1,643 sq. deg. & 1,253 sq. deg. & $<$ 50 sq. deg. & 506 sq. deg. & 307 sq.deg.\\
\rule[-1ex]{0pt}{3.5ex} Redundant pixels & 15.36\% & 14.8\% & $< 1\%$ & 5.2\% & 5.5\% \\
\hline
\end{tabular}
\end{center}
 \vspace{-0.25cm}\hspace{0.5cm} a: Tessellation frequency of the underlying triangular subdivision. Subdivision scheme is Equal-arcs (three great circles). b: average pixel size including gaps between pixels. c: at $\lambda=0.5\mu m$.
\end{table} 

\section{CONCLUSIONS AND FURTHER WORK} \label{sec:conclusion}

Presented is a novel preliminary design for a panoramic optical pulsed SETI observatory, for which
we propose to use two geodesic assemblies of 198 Fresnel telescopes to search for technosignatures in the visible and near-infrared with low angular resolution ($>$2-4 arcmin) and $\sim 10,000$sq-deg instantaneous FoV.
Validation of this design through detailed structural analysis will allow determination of key structural parameters.
Refractive Fresnel lenses are ideally suited for low ($>2$arcmin) angular resolution observations, with large pixels ($>$0.3mm) operating as light buckets. Achromatization of the optical system will be investigated for larger bandwidth observations. 

\acknowledgments  
 
The PANOSETI research and instrumentation program is made possible by the enthusiastic support and interest by Franklin Antonio. We thank the Bloomfield Family Foundation for supporting SETI research at UC San Diego in the CASS Optical and Infrared Laboratory. Harvard SETI is supported by The Planetary Society. UC Berkeley's SETI efforts involved with PANOSETI are supported by NSF grant 1407804, the Breakthrough Prize Foundation, and the Marilyn and Watson Alberts SETI Chair fund. Lastly, we would like to thank the staff at Mt. Laguna and Lick Observatories for their help with equipment testing.

\bibliography{report}
\bibliographystyle{spiebib}

\end{document}